\documentclass[12pt,journal,comsoc]{IEEEtran}
\ifCLASSOPTIONcompsoc
  \usepackage[nocompress]{cite}
\else
  \usepackage{cite}
\fi
\ifCLASSINFOpdf
  \usepackage[pdftex]{graphicx}
  \graphicspath{{../figures/}}
  \DeclareGraphicsExtensions{.pdf,.jpeg,.png}
\else
\fi
\usepackage{xcolor}
\usepackage{soul}
\usepackage{multirow}

\begin{document}
%
\title{Watching Smartly from the Bottom: Intrusion Detection revamped through Programmable Networks and Artificial Intelligence}
%

%
%

\author{Sergio~Armando~Gutiérrez,
        John~Willian~Branch,
        Luciano~Paschoal~Gaspary,
        Juan~Felipe~Botero
\IEEEcompsocitemizethanks{\IEEEcompsocthanksitem S.A. Gutiérrez is Researcher and Lecturer at UNAULA, Medellín, Colombia.\protect\\
E-mail: sergio.gutierrezbe@unaula.edu.co
\IEEEcompsocthanksitem J.W. Branch is Full Professor at UNAL, Medellín, Colombia.\protect\\
E-mail: jwbranch@unal.edu.co
\IEEEcompsocthanksitem L. Gaspary is Associate Professor at UFRGS, Porto Alegre, Brazil.\protect\\
E-mail: paschoal@ufrgs.br
\IEEEcompsocthanksitem J.F.Botero is Associate Professor at UDEA, Medellín, Colombia.\protect\\
E-mail: juanf.botero@udea.edu.co

}
\thanks{Manuscript received October 1, 2020; revised October 1, 2020.}}

%
%

\def\href#1#2{#2}

\IEEEtitleabstractindextext{%
\begin{abstract}
A recent research line has explored the possibility of leveraging functionalities of Programmable Data Planes to offload part of Machine Learning algorithms to the data plane, which might contribute to increase their accuracy and responsiveness by having a more detailed visibility of the traffic. This approach introduces a significant opportunity for evolution in the critical field of Intrusion Detection. In this paper, we discuss how Programmable Data Planes might complement different stages of an Intrusion Detection System based on Machine Learning. We present two use cases that make evident the feasibility of this approach and highlight aspects that must be considered when addressing the non straightforward task  of deploying solutions leveraging data-plane functionalities. 
\end{abstract}

}

\maketitle

\IEEEdisplaynontitleabstractindextext

%
\IEEEpeerreviewmaketitle

\section{Introduction}\label{sec:introduction}


\IEEEPARstart{T}{he} emergence of the Programmable Data Planes (PDPs) concept has represented an outstanding advance in the complete realization of the Software Defined Networking (SDN) Paradigm. The capacity of influencing Forwarding Device (FD) behavior via abstractions such as programming languages is an idea that has been explored through different initiatives proposed by industry and academia. An important milestone in the evolution of network programmability was the development of OpenFlow as an implementation of SDN. However in traditional SDN, the network operation is still constrained to the set of protocols and headers supported by the hardware of the FDs. Therefore, the definition of custom functions for packet processing becomes very difficult.

Recently, the concept of PDPs has emerged to overcome these constraints by allowing going beyond the definition of flow rules. PDPs enable a complete control of the network behavior, from the applications to the packet processing within the FDs, including the definition and parsing/deparsing of custom headers. It provides an unprecedented opportunity to develop new features within the FDs and revisiting existing functions for network management~\cite{da2017data}. P4 is currently the \textit{de-facto} standard language for describing how network packets should be processed.

One of the areas that can take advantage of PDP/P4 is network security. Particularly, Intrusion Detection Systems (IDSs) can be revamped by implementing them as efficient functions deployed at the data plane, capable to react promptly to network anomalies that might represent threats. IDSs typically depend on the collection of traffic features which are later fed into sophisticated systems mostly based on Machine Learning (ML) algorithms. ML has been successfully used in network security due to its capacity to detect and discover patterns and behaviors previously unseen in the network traffic. Most of the approaches for security developed in the context of SDN and based on ML have been implemented at the control plane despite the issues associated to accuracy and high overhead they might introduce~\cite{xie2018survey}.

The features introduced by PDPs make possible to consider a new landscape for security solutions based on ML, by conceiving offloading part of the algorithms to the FDs. Thus, more accurate and responsive solutions can be deployed. Delegating part of the functions to the data plane might also help to reduce the overhead at the control plane. Nevertheless, the decision regarding how much of the functions should be offloaded to the data plane is not trivial~\cite{ports2019should} because the computation capabilities of FDs are limited, and an excessive offloading of functionalities might hinder the maximum throughput in packet forwarding. 

In this paper, \textcolor{black}{our main contribution is presenting} a critical discussion on the challenges introduced when intersecting Programmable Data Planes and Artificial Intelligence (AI) for intrusion detection. We focus on devising how to leverage the functionalities offered by PDPs in the implementation of AI and specially ML algorithms. We specially try to address the question regarding how much of the operations of the algorithms is feasible to be offloaded to the FDs. 

\section{Machine Learning at the Programmable Data Plane for Improved Intrusion Detection}
\label{sec:sectionStages}

\begin{figure*}[!t]%
    \centering
    {\includegraphics[scale=0.65]{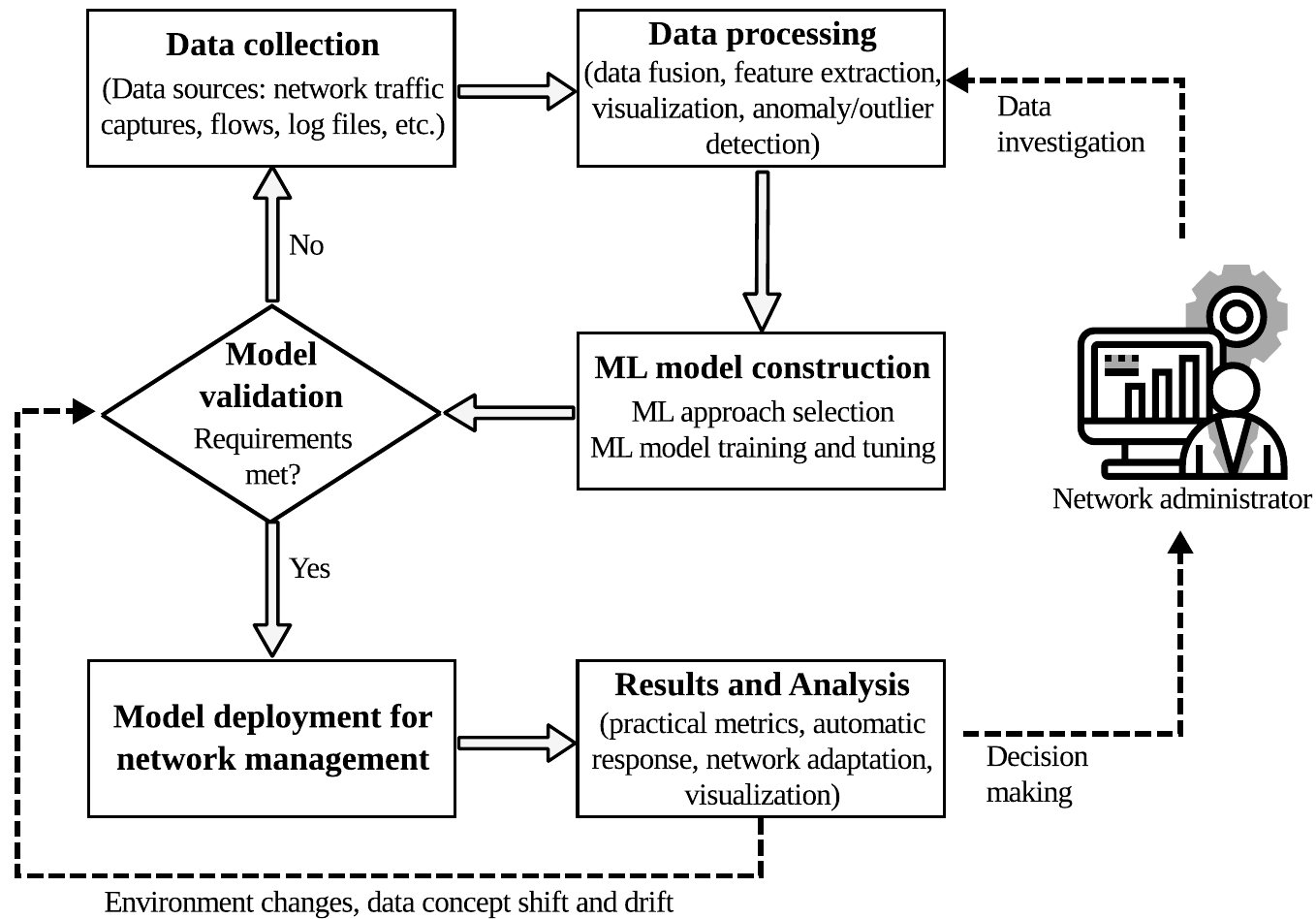} }%
    \caption{Stages of a ML system applied to Network Management domains (e.g. Intrusion Detection)~\cite{le2020frontier}}
    \qquad
    \label{fig:stages}%
\end{figure*}

ML has become an important milestone in several kinds of cyber security solutions due to its ability to extract, in a timely fashion, anomalies and patterns that might be symptom of internal or external attacks against the infrastructure. These solutions are usually integrated by network and host security components, minimally including firewalls, anti viruses and Intrusion Detection Systems (IDSs)~\cite{le2020frontier}.

IDSs are being actively revisited in the context of programmable networks. Most of the solutions developed for implementing IDSs have been deployed as applications running in the control plane. However, this approach for the implementation of IDSs has two main drawbacks. First, the set of traffic features derived either from standard counters (\textit{e.g.}, those available in current versions of OpenFlow) or, in extreme situations, via the \texttt{PACKET\_IN} events, are insufficient to get reasonable accuracy in ML algorithms. Second, ML algorithms are typically compute-intensive tasks, which, if not properly designed/deployed, might introduce control plane overhead and disrupt the correct operation of the network~\cite{binbusayyis2019identifying}.

As we previously introduced, the advent of PDP makes possible the offloading of some functionalities to the data plane. This is an strategy that might help to overcome the limitations associated to scarce information as input while reducing the overhead of the control plane. Next, we present an overview of the stages of an ML-based IDS, and we outline how PDPs can be leveraged to improve some of these stages through custom packet processing and the offloading of certain operations~\cite{le2020frontier}. Figure \ref{fig:stages} presents the sequential view and relationships of these typical stages.

\textbf{Data Collection.} PDPs extend the possibilities for data collection beyond the standard statistics available in the FDs. Custom statistics can be introduced, and some stateful processing can be included within FDs, which might even indicate events with high value for intrusion detection tasks~\cite{kohler2018p4cep}. 
Despite the limitations inherent to the computing power and programming primitives available in programmable FDs, there are two functionalities that can be leveraged to implement efficient data collection for ML algorithms: Custom packet parsing and data aggregation. Custom Packet parsing allows processing headers that can be used to store personalized information and statistics. These headers enable differentiated processing and information gathering within FDs, or applying custom hash functions in order to distinguish specifics sets of packets~\cite{gupta2018sonata}. On the other hand, data aggregation helps to reduce the amount of data that needs to be transmitted in the network from FDs to Control Plane, for the execution of complex operations there~\cite{sapio2019scaling}. 

\textbf{Data Processing.} Recent proposals introduce the concept of In-Network processing as a service, where the implementation of the mentioned operations in data plane becomes available to be used by high level ML or data-analytics frameworks~\cite{mustard2019jumpgate, sapio2017network}. Other proposals introduce the notion of queries that trigger the joint collection and preliminary analysis of data in order to produce statistics that can be later delivered to higher level processing mechanisms~\cite{gupta2018sonata}. The main notion behind these approaches is the fact that features of PDPs such as programmable parsing, state storing, custom hashing and the flexibility of match/action tables allow not only measurements and counts, but also performing analysis in parallel to the data collection. 

\textbf{Model Construction.} The literature presents several approaches to leverage the features available in PDP for the implementation of different algorithms considering the computational constraints of programmable FDs~\cite{ports2019should}. These proposals include the use of switch registers to implement bit-based value storage and arithmetic which might be even useful to implement Neural Networks~\cite{qin2020line} and match action and long prefix match tables with comparison operators for the implementation of algorithms such as decision trees, support vector machines and Naive Bayes classifiers. This approach reduces the amount information that needs to be forwarded towards control plane (e.g. \texttt{PACKET\_IN} events) which contributes to diminish the control channel overhead~\cite{macas2020oracle} while increases the accuracy and responsiveness~\cite{xiong2019switches, ports2019should}.
In addition to the direct implementation within the FDs, another approach to be followed is cooperating in the training of large scale models via local analysis of metrics. This approach is called Federated Learning, and can be leveraged for the training of complex models such as Deep Neural Networks~\cite{qin2020line}.

\textbf{Model Validation.} Validation is a high level task which involves extensive analysis and feedback provided by human experts. Hence, PDPs do not have a direct intervention in the tasks associated to this stage. However, functionalities such as Complex Event Processing~\cite{kohler2018p4cep} and query-driven telemetry~\cite{gupta2018sonata} are helpful to provide insights to debug situations of low accuracy and poor performance of the ML algorithms.

\textbf{Deployment.} This operation must consider the particularities involved in the development of the algorithms. For example, the availability in a given hardware architecture of the type of tables required, or the amount of registers that may be used to store packet state are aspects that must be validated~\cite{xiong2019switches,qin2020line}. For a detailed discussion on the issues associated to the deployment of ML algorithms within FDs, please refer to~\cite{xiong2019switches} and~\cite{ports2019should}.

\textbf{Result Analysis.} Functionalities such as In-Band Network Telemetry, which relies on PDP features~\cite{gupta2018sonata}, and Complex Event Processing~\cite{kohler2018p4cep} might provide important insights for this stage. Also, the definition of thresholds for specific features and time intervals~\cite{lapolli2019offloading}, and even the support for query-driven information collection might contribute to assess the effectiveness of the algorithms while allowing some degree of data analysis.

Table \ref{table:1} summarizes our insights regarding where to implement the particular functions, either in the control or the data plane and the Machine Learning Algorithm(s) that might be involved.


\begin{table*}[tp]
\renewcommand{\arraystretch}{1.3}
\caption{Place of deployment for functionalities associated to an ML-Based IDS}
\label{table:1}
\centering
\begin{tabular}{l|c|c|c|c}
Stage & Functionality & Where to implement & ML algorithm & Related Work \\
\hline
Data Collection & Custom statistics and metrics & Data Plane & CUSUM & \cite{kohler2018p4cep} \\ 
Data Collection & Header hashing & Data Plane & Sketches & \cite{gupta2018sonata} \\ 
Data Collection & Custom packet parsing & Data Plane & KNN & \cite{mustard2019jumpgate,sapio2017network} \\
Data Processing & Partial state storing & Data Plane & Sketches & \cite{xiong2019switches} \\
Data Processing & Threshold comparison & Data Plane & CUSUM,SVM & \cite{lapolli2019offloading} \\
Data Processing & In-network processing as a Service & Control Plane & SVM,KNN,PCA & \cite{mustard2019jumpgate}     \\
Model Construction & Complex Arithmetic & Control Plane & NN,SVM,RNN & \cite{sapio2019scaling,lewis2019p4id,qin2020line}\\
Model Validation & Complex Event Processing & Control Plane & RNN & \cite{kohler2018p4cep} \\
Model Validation & Query-Driven Telemetry & Control Plane & Sketches,RNN & \cite{gupta2018sonata}  \\
\hline
\end{tabular}
\end{table*}

The processing capabilities available nowadays in programmable devices are mainly useful for data collection and data processing. Currently, most of the functionalities associated to ML algorithms such as Cumulative Summation Chart Control (CUSUM), Support Vector Machines (SVM) K-Nearest Neighbors (KNN), Principal Component Analysis (PCA), Basic Neural Networks (NN) or Recurrent Neural Networks (RNN) tend to be implemented within the control plane. However, the implementation of ML algorithms within the forwarding devices can be expected in a near future, by taking advantage of more robust hardware platforms for these devices.

\section{Use Cases}
\label{sec:useCases}

In the previous section, we presented an overview about how PDPs can contribute in the implementation of some of the stages of a system based on ML algorithms applied to the domain of IDS. In this section, we present two use cases which exemplify how to specifically leverage functionalities of PDPs in the context of security solutions based on these ML algorithms.

\subsection{Offloading Data Collection and Processing in DDoS attack Detection to Programmable Data Planes}

DDoS attack detection in traditional SDN environments introduces two design challenges: 1) the control plane is overloaded due to the continuous polling of network traffic information as well as the calculation and classification of the feature set; and 2) OpenFlow cannot obtain specific features fundamental for DDoS detection (e.g. per-flow packet inter-arrival times and packet payload sizes). These two challenges can be overcome by offloading some of the ML-based detection stages from the control plane to P4-based PDP.

\begin{figure}[t!]
    \centering
    \includegraphics[scale=0.6]{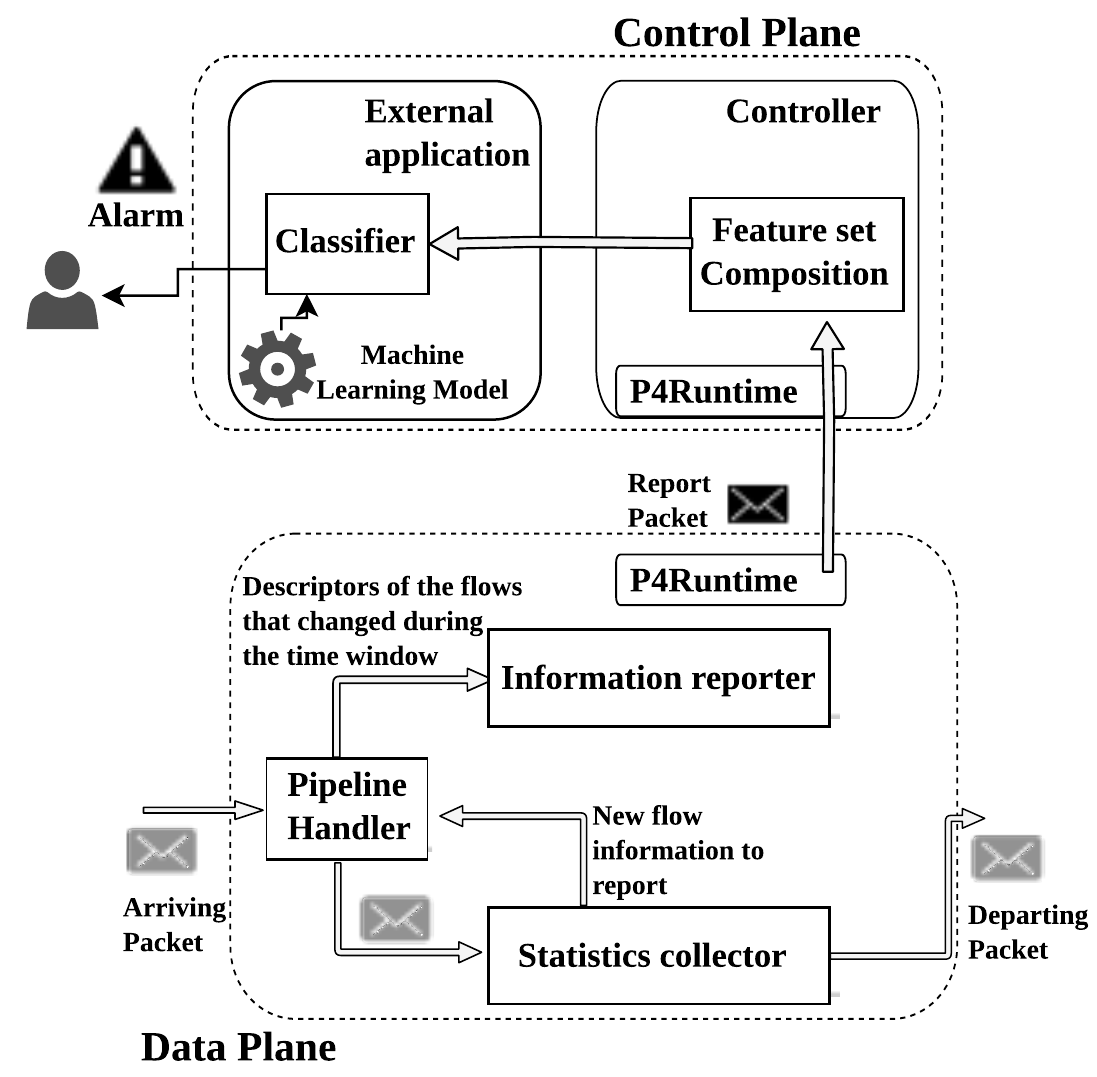}
    \caption{ML-based DDoS attack detection architecture coordinating control and programmable data planes~\cite{macas2020oracle}}
    \label{fig:Arch}
\end{figure}

Fig.~\ref{fig:Arch} shows a DDoS detection network architecture where some ML stages are offloaded from the \textit{control plane} to the \textit{programmable data plane} that implements a time window-based scheme. During each window, it simultaneously: i) collects new flow information; i.e. it performs the \textit{Data Collection} stage mentioned in Section~\ref{sec:sectionStages} by extracting specific per-flow information needed to build the feature set for the attack detection, and ii) reports the information collected in the previous window to the control plane; i.e. it carries out the \textit{Data Processing} stage by pre-processing the needed per-flow features and forwarding them to the control plane that, in turn, easily generates the feature set to feed the ML-based detection algorithm.

When a new packet arrives to the FD, the \textit{statistics collector} extracts the packet information building specific flow descriptors using just the arithmetic operations available in P4. Then, these descriptors are sent to the control plane for the generation of the feature set (detailed information of the descriptors is provided in~\cite{macas2020oracle}). Once the time window expires, the \textit{pipeline handler} checks if the \textit{statistics collector} contains new flow information to report and, if this is the case, the information is forwarded to the \textit{information reporter} which generates a report packet with the specific descriptors of the flows that changed during the time window. Finally, the report packet is forwarded to the control plane using the \texttt{P4Runtime} interface.

The control plane is composed of an SDN controller connected to a ML-based traffic classification application. The \textit{feature set composition} component parses the custom header of the packet that arrived from the data plane and builds a tuple with the resulting per-flow feature set. Finally, this set is received by the \textit{classifier} application that executes an already loaded ML model which performs binary classification (i.e. determines if each received tuple belongs either to \textit{DDoS} class or \textit{benign} class).

In this approach, we implemented two ML classifiers for different time window duration: Random Forest (RF) and K-Nearest Neighbor (KNN). The experimental evaluation consisted on reproducing a DDoS attack contained on a state-of-the-art dataset. The experimental environment was built on mininet, using the reference P4 Switch implementation (BMV2) \cite{da2017data}. The comparison of the two classifiers used showed that KNN achieved better performance than RF. Obtained results led to an accuracy from 93\% to 96\% for the different time windows, which is within the upper range seen in the literature. 

Besides reaching high accuracy, the offloading of the data collection and processing stages to the PDP provides two main advantages. 1) A configurable time window forwarding just one report packet to the control plane removes the need of duplicating packets and send them to external IDS. It also prevents the continuous transfer of OpenFlow statistics from the FD to the control plane, therefore reducing traffic overload in the control channel. 2) Data plane programmability allows to extract just the feature descriptors needed by the control plane to build the feature set in order to perform the ML-based DDoS attack detection. In contrast, traditional SDN-OpenFlow collects statistics that are useless to detect DDoS attacks. Besides, OpenFlow is unable to collect some features that are paramount to classify DDoS attacks with ML-based algorithms (please see~\cite{macas2020oracle} for further details). 

\subsection{Offloading Real-Time DDoS Attack Detection to Programmable Data Planes}

There are situations (\textit{e.g.}, in service provider core networks) in which the fast, real-time detection of an intrusion is a primary requirement, even if sacrificing (a little of the) accuracy. Next, we briefly present a mechanism in which we devise a lightweight entropy-based anomaly detection technique and offload it \textit{entirely} to a programmable forwarding device based on P4~\cite{lapolli2019offloading}.

\begin{figure}
    \centering
    \includegraphics[scale=0.7]{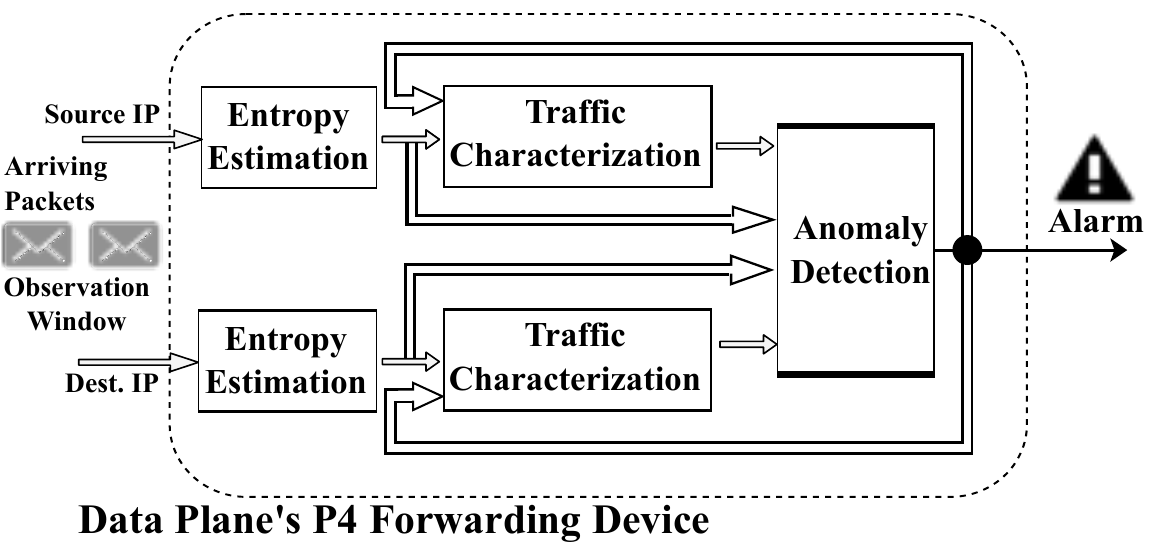}
    \caption{In-network mechanism for real-time DDoS attack detection~\cite{lapolli2019offloading}}
    \label{fig:Mech}
\end{figure}

Figure \ref{fig:Mech} illustrates the main architectural components of the mechanism. It implements a processing pipeline to calculate the entropies of source and destination IP addresses of packets arriving at an FD. They are estimated for consecutive packets of the incoming stream, named \textit{observation windows}. Upon completing each window, the traffic characterization components (for source and destination IPs) consume the entropy values to produce a legitimate traffic model. The anomaly detection component calculates detection thresholds as functions of this model, issuing an attack alarm when the last entropy estimates exceed them.

Given the restrictions of the primitive set of operations provided by P4, a simplified design was compulsory in several circumstances. For example, we employed customized count sketches to approximate the frequencies of different IP addresses. To solve compute-intensive arithmetic functions, we used longest-prefix match (LPM) lookup tables. The lack of procedures to iterate over data structures required the design of incremental, per-packet computation methods. Such adjustments played a vital role in enabling the offload of several ML system stages to the data plane (see Figure \ref{fig:stages}), such as data collection and processing (entropy values), model construction (legitimate traffic), and analysis (threshold-based alarming).

\textcolor{black}{A set of experiments for assessing the efficacy and efficiency of our proposal was performed on a software-based P4 infrastructure. These experiments used datasets from CAIDA representing normal and DDoS traffic. The DDoS traffic data set is built in such a way that it contains only attack-related traces, and has been used in several high impact papers. Regarding the use of a software-based infrastructure, we consider that it does not limit our assessment, since accuracy and resource utilization are expected to be equivalent regardless the actual hardware target.}

For low-intensity attacks, \textit{e.g.}, occupying a fraction of 4\% of the whole traffic, the detection accuracy exceeds 90\%. Our approach requires a few hundred milliseconds to detect attacks since it perform per-packet analysis (as opposed to sampling). We refer the reader to~\cite{lapolli2019offloading} for a detailed description of this work.

\section{Discussion and insights}
\label{sec:discussion}

The implementations described above are by themselves evidence on the feasibility of achieving the effective combination of ML and PDPs for network security solutions. However, from the analysis of these possibilities, the aforementioned use cases, and considering the examples of solutions available in the literature, there are several points that we would like to highlight since they introduce important and relevant insights leading to further research lines.

\textbf{Executing Complex Operations on the Data Plane.} The set of mathematical operations supported by programmable switches is limited. While implementing addition or logical bit shift operations is easy, general mathematical operations such as those commonly involved in ML algorithms (\textit{e.g.}, square roots, divisions and in general floating point arithmetic) can not be implemented. An approach proposed in the literature to overcome this situation is using the tables within the switch to store values computed externally, corresponding to hyperparameters of the ML algorithms upon their training. Thus, the part of the algorithms running at the FDs consists of matching, conditional and logic operators, which are naturally available within the devices~\cite{xiong2019switches, lapolli2019offloading}.

\textbf{Lightweight ML Algorithms in PDPs.} Due to resource limitation, the algorithms to be implemented within the FDs need to be lightweight. An strategy to simplify the algorithms is the one previously mentioned of using look up tables. The tables would contain the results of complex calculations that are performed outside the FDs (\textit{e.g.}, at the control plane). This could be combined with additional strategies such as the simplification of data representation for the ML algorithms. Binarized Neural Networks are an example, in which the weights associated to the Neural Network are replaced by binary bits stored within registers in the FDs~\cite{qin2020line}.  

\textbf{Centralized vs. Distributed PDP-based ML Systems.} A centralized operation would be the trivial scenario by taking advantage of the local traffic visibility. However, for complex ML algorithms (\textit{e.g.}, Deep Neural Networks), this would introduce an important overhead in operations such as feature extraction. An approach to overcome this overhead is \textit{Federated Learning}. In this approach, FDs have a copy of the trained ML model, and improve them according to their local observation. The adjusted model is sent back to the control plane, where it is combined with the information provided by other FDs, and it is further distributed to all the devices~\cite{qin2020line}.

\textbf{Scope of decisions according to the amount of information.} PDPs enable data collection in local contexts via the observation of traffic crossing the forwarding device. This capability allows quick reaction upon the detection of anomalies or shifts in pattern traffics which might be symptom of threats or ongoing attacks. However, due to the locality of this observation, the collected information can not be leveraged for strategic decisions by itself. This information needs to be collected and somehow integrated in more comprehensive analysis in order to make strategic decisions based on a global view of the acquired information \cite{sapio2019scaling,qin2020line}.

\textbf{What to Execute at the Data and Control Planes?}
The literature has shown two main different approaches where the functionalities of programmable FDs make feasible the implementation of simple and even complex ML algorithms. The first approach mainly consists in implementing simple tasks such as counting, classification, aggregation and grouping. Depending on the complexity of the primitive operations involved (e.g. arithmetic), the implementation might be complete or partially supported with the external execution of those operations not available within the FD~\cite{xiong2019switches,qin2020line,macas2020oracle,lapolli2019offloading}. The second approach consists in offering the operations available in the FD as services which can be invoked from the Control Plane. Hence, these operations can be leveraged from complex algorithms, offloading the low level and simpler tasks to the FD while leaving the more complex operations to be performed at control plane. This approach reduces the overhead by avoiding excessive information exchange between control and data plane~\cite{gupta2018sonata,mustard2019jumpgate}. 


\section{Conclusions}
\label{sec:conclusions}

In this paper, we have discussed the contributions that PDPs bring to the development of IDS solutions based on ML algorithms. \textcolor{black}{In the presented use cases we showed the implementation of traditional ML algorithms such as KNN. However, more complex and state-of-the-art solutions such as Deep Neural Networks can also leverage PDPs. Initial efforts on this trend can be observed on the interest of this topic evident in different high quality research venues such as the NETAI Workshop which is part of the SIGCOMM conference, the ACM HotNETS workshop, and IEEE IM Conference, among others.} Although data-plane functionalities provide more sophisticated possibilities, implementing functions at forwarding devices is a non-trivial task that needs to be carefully performed to explore the advantages while preserving high performance in packet forwarding.

\ifCLASSOPTIONcompsoc
  \section*{Acknowledgments}
\else
  \section*{Acknowledgment}
\fi

This paper has been supported by the project ``Red Temática
CYTED 519RT0580'' funded by the Ibero-American Science and Technology Program CYTED.

\ifCLASSOPTIONcaptionsoff
  \newpage
\fi



\bibliographystyle{IEEEtran}
\bibliography{IEEEabrv,./bibliography.bib}
%



%
\vskip -2.5\baselineskip plus -1fil

\begin{IEEEbiographynophoto}{Sergio Armando Guti\'errez} Received his PhD in Computer Science from Universidad Nacional de Colombia, Medellín in 2018. He is currently full time lecturer at Universidad Autónoma Latinoamericana. His research interests include computer networks, security in computer networks, , programmable devices and application of machine learning to computer networks. ORCID: 0000-0003-2880-4601
\end{IEEEbiographynophoto}
\vskip -3\baselineskip plus -1fil
\begin{IEEEbiographynophoto}{Luciano Paschoal Gaspary} received his Ph.D. in Computer Science from Universidade Federal do Rio Grande do Sul, Brazil in 2002. He is currently Deputy Dean and Associate Professor at the Institute of Informatics at UFRGS. His research interests include computer networks, network management, and computer system security. ORCID: 0000-0002-7561-5582
\end{IEEEbiographynophoto}
\vskip -3\baselineskip plus -1fil
\begin{IEEEbiographynophoto}{Juan Felipe Botero} received his Ph.D. degree in Telematics engineering from Universidad Técnica de Cataluña, Spain, in 2013. He is an Associated Professor with the Electronics and Telecommunications Engineering Department at Universidad de Antioquia, Colombia. His main research interests include quality of service, software defined networking, network virtualization, and resource allocation. ORCID: 0000-0002-7072-8924
\end{IEEEbiographynophoto}
\vskip -3\baselineskip plus -1fil
\begin{IEEEbiographynophoto}{John Willian Branch} received his PhD in Systems Engineering in 2007 from Universidad Nacional de Colombia, Medellín. He is a full professor at the Computing and Decision Sciences Department, in the Facultad de Minas, Universidad Nacional de Colombia, Medellin, Colombia. His research interests include: automation, computer vision, digital image processing and computational intelligence techniques applied in different domains ORCID: 0000-0002-0378-028
\end{IEEEbiographynophoto}




\end{document}